\documentclass[
    ,final            
  ]
  {aipproc}

\layoutstyle{6x9}

\begin{document}

\title{Relic abundance and detection prospects of neutralino dark
      matter in mirage mediation
}


\author{Ken-ichi Okumura
\footnote{okumura@higgs.phys.kyushu-u.ac.jp}
\footnote{This talk is based on the collaboration in \cite{Choi:2006im}.}
}{
  address={Department of Physics, Kyushu University, Fukuoka 812-8581, Japan},
}

\begin{abstract}
We analyze thermal relic
 abundance of neutralino dark matter and discuss future prospects of its direct
 and indirect detections in mirage mediation.
\end{abstract}

\maketitle


\section{Introduction}
Supersymmetry (SUSY) is a promising candidate for physics beyond the standard
model (SM) which provides an elegant solution of the gauge hierarchy
problem and is also endowed with various phenomenological virtues,
 including renowned gauge coupling unification.
SUSY with R parity naturally introduces 
the lightest supersymmetric particle (LSP) as a candidate for cold dark
matter.
If gravitino is heavy enough, the best candidate is LSP neutralino.
 On the other hand precision cosmology in these decades has pinned down
 its abundance in the universe. 
This enables us to select realistic SUSY models via the nature of LSP
 and calculation of its relic abundance under plausible assumption on the
 evolution of the universe.
This is particularly interesting in view of making contact between 
new physics search in LHC and direct/indirect detection of cold dark matter.

Recently, a new class of SUSY breaking scenario called mirage mediation
 has been proposed \cite{Choi:2004sx, Loaiza-Brito:2005fa}, which is also
 known as the mixed modulus-anomaly mediation. It is
 a generic consequence of the KKLT-type moduli stabilization
 \cite{Kachru:2003aw} and its low energy SUSY
 spectrum is distinct from those of the conventional SUSY breaking scenarios
 such as minimal supergravity (mSUGRA), gauge mediation and anomaly
 mediation \cite{Choi:2005uz,Endo:2005uy}.
Phenomenological and cosmological aspects of it have been investigated by
 several groups \cite{Choi:2005uz,Endo:2005uy,falkowski05,yama}.
%
In this paper, we summarize the results of our analysis on
 the thermal relic abundance of neutralino dark matter
 in mirage mediation and discuss
 prospects of its direct and indirect detections in  ongoing and planned 
 experiments.

\section{Mirage mediation}
Here we briefly summarize the mirage mediation. More detailed explanations
 are found e.g. in \cite{Choi:2005uz, Choi:2006im}.
Mirage mediation is a natural consequence of the following two
 assumptions on moduli stabilization and SUSY breaking:
\begin{itemize}
\item {Gauge coupling modulus $T$ for visible gauge fields is stabilized
      by non-perturbative dynamics which does not break SUSY itself.}
\item{SUSY is broken by a brane-localized source which is sequestered
      from visible sector.}
\end{itemize}
Both of them are typically realized in KKLT moduli stabilization in type
IIB string \cite{Kachru:2003aw}. 
Small shift of the modulus vev by SUSY breaking uplifting
 leads to suppressed F-component relative to the gravitino mass \cite{Choi:2004sx},
\begin{eqnarray}
M_0 \equiv F^T/(T+T^\ast) \approx m_{3/2}/\ln\left(M_{Pl}/m_{3/2}\right),
\end{eqnarray}
which gives mixed modulus-anomaly mediated patterns to the soft SUSY
breaking terms at the unification scale, $M_{GUT}$,
\begin{eqnarray}
&&{\cal L}_{\rm
soft}=-\frac{1}{2}M_a\lambda^a\lambda^a-\frac{1}{2}m_i^2|\phi_i|^2
-\frac{1}{6}A_{ijk}y_{ijk}\phi_i\phi_j\phi_k+{\rm h.c.}, \nonumber\\
&&M_a = M_0 + M_a^{(3/2)},
~~m^2_i = \widetilde{m}^2_i + m^{2 (3/2)}_i + \Delta m^{2}_i,
~~A_{ijk} = \widetilde{A}_i + \widetilde{A}_j + \widetilde{A}_k + A_{ijk}^{(3/2)},
\end{eqnarray}
where $M_0$ and tilde indicate modulus mediation and superscript $(3/2)$ denotes anomaly mediation.
$\Delta m^2_i$ represents their interference.
Note $\ln\left(M_{Pl}/m_{3/2}\right) \approx 4\pi^2$ and the two
contributions are comparable.
We introduce the following dimensionless parameters,
\begin{eqnarray}
\alpha\,\equiv\,m_{3/2}/M_0\ln(M_{Pl}/m_{3/2}),\quad
 a_i\,\equiv\,\tilde{A}_{i}/M_0, \quad
 c_i\,\equiv\,\tilde{m}_i^2/M_0^2.
\label{eq:def}
\end{eqnarray}
$\alpha$ is dynamically set real due to shift symmetry of $T$
 and R symmetry \cite{susycp}.
$a_i$ and $c_i$ are typically non-negative rational numbers,
 depending upon scaling property of the matter K\"ahler metric against
 $T+T^\ast$.  
In toroidal compactification, they are given by $a_i = c_i = 1 -n_i$ 
 for matter fields with modular weight $n_i$.
A unique feature of the mirage mediation is the fact that the addition of anomaly mediation
 is equivalent to effective shift of the modulus mediation scale
 from $M_{GUT}$ to
 {\it the mirage messenger scale}, $M_{\rm
 mir}=M_{GUT}/(M_{Pl}/m_{3/2})^{\alpha/2}$ \cite{Choi:2005uz}. 
This is true not only for gaugino masses but also for trilinear couplings and soft scalar masses if relevant Yukawa couplings are negligible or only allowed for combination
 of $a_i+a_j+a_k=c_i+c_j+c_k=1$. 
Note that the unification scale of the gauge couplings itself remains
 the same. 
This leads to rather degenerate
 low energy spectrum distinct from
 mSUGRA, gauge mediation or anomaly mediation 
 and consequently its characteristic phenomenology.

In the case of neutralino dark matter, reduced gluino/bino mass ratio,
\begin{eqnarray}
M_3:M_2:M_1\simeq
(1-0.3\alpha)g_3^2:(1+0.1\alpha)g_2^2:(1+0.66\alpha)g_1^2,
\end{eqnarray}
results in reduced higgsino and heavy Higgs boson masses relative to
 the bino mass through the RG running under constraint of radiative
 electroweak (EW) symmetry breaking.
This enhances higgsino components in the lightest neutralino
 and also increases a chance for neutralino to annihilate via heavy Higgs
 boson resonance.

If SUSY breaking sector is sequestered by warped throat as in the KKLT moduli
 stabilization and gauge couplings are determined only by $T$,
 $\alpha = 1$ and $M_{\rm mir} \sim 3\times 10^{9}$ GeV are predicted.
However, if we introduce dilaton-modulus mixing in gauge kinetic
 functions for visible gauge fields and/or non-perturbative dynamics
 of the modulus stabilization, the more varied value of $\alpha$ is
 available \cite{abe}.
In the next section we first examine the minimal case $\alpha=1$
 and later extend it for general cases.

\section{Relic abundance and detection prospects}

In this section, we analyze the thermal relic abundance of the
neutralino dark matter in mirage mediation and prospects of its direct and
indirect detection.
We assume that the universe undergoes
 standard thermal evolution during and after the decoupling of
 neutralino which occurs at a temperature, $T_{\chi} \sim m_{\chi}/20$
where $m_\chi$ is the lightest neutralino mass.
For the numerical calculation of the abundance and various observables
in direct/indirect detection,
 we extensively use the {\it Dark SUSY 4.1} package \cite{darksusy}. 

The mirage mediation inherently involves a light modulus of
 $m_{T}\sim4\pi^2 m_{3/2}$.
After the inflation, it is known that if displaced such a modulus starts
 coherent oscillation and eventually
 dominates the energy density of the universe.
Its subsequent decays to gravitino and other MSSM fields
 reheat the universe again.
Recently, this process is examined in detail and 
 non-thermal neutralino from the decay of the gravitino
 is found to overclose the universe \cite{yama}.
We assume this non-thermal contribution is diluted 
 before the universe reaches $T_{\chi}$. 
 Such a dilution is possible by thermal inflation \cite{stewart}.

\begin{figure}
\begin{tabular}{cc}
  \scalebox{1.}{\includegraphics[height=.24\textheight]{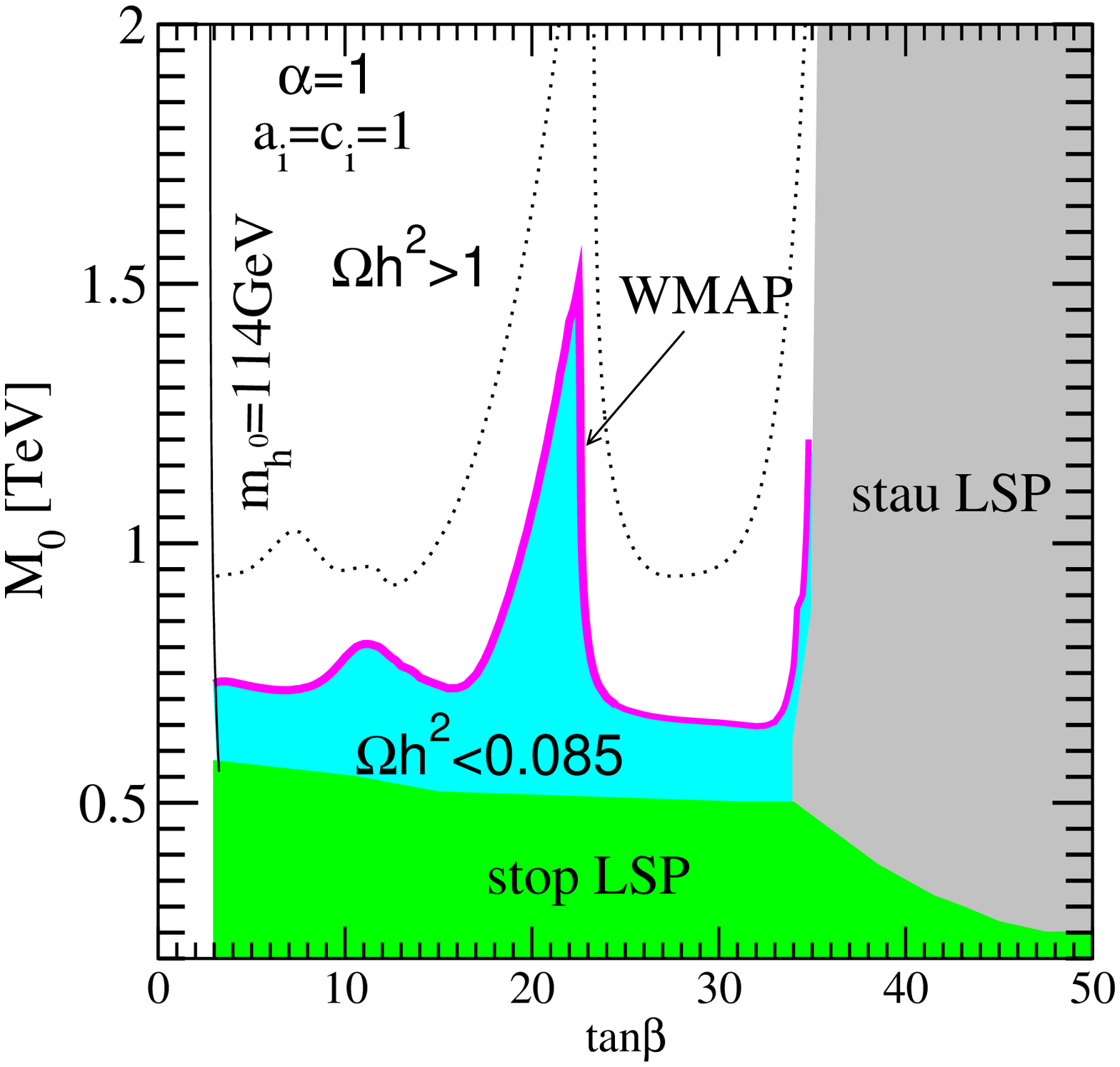}} &
  \scalebox{1.}{\includegraphics[height=.24\textheight]{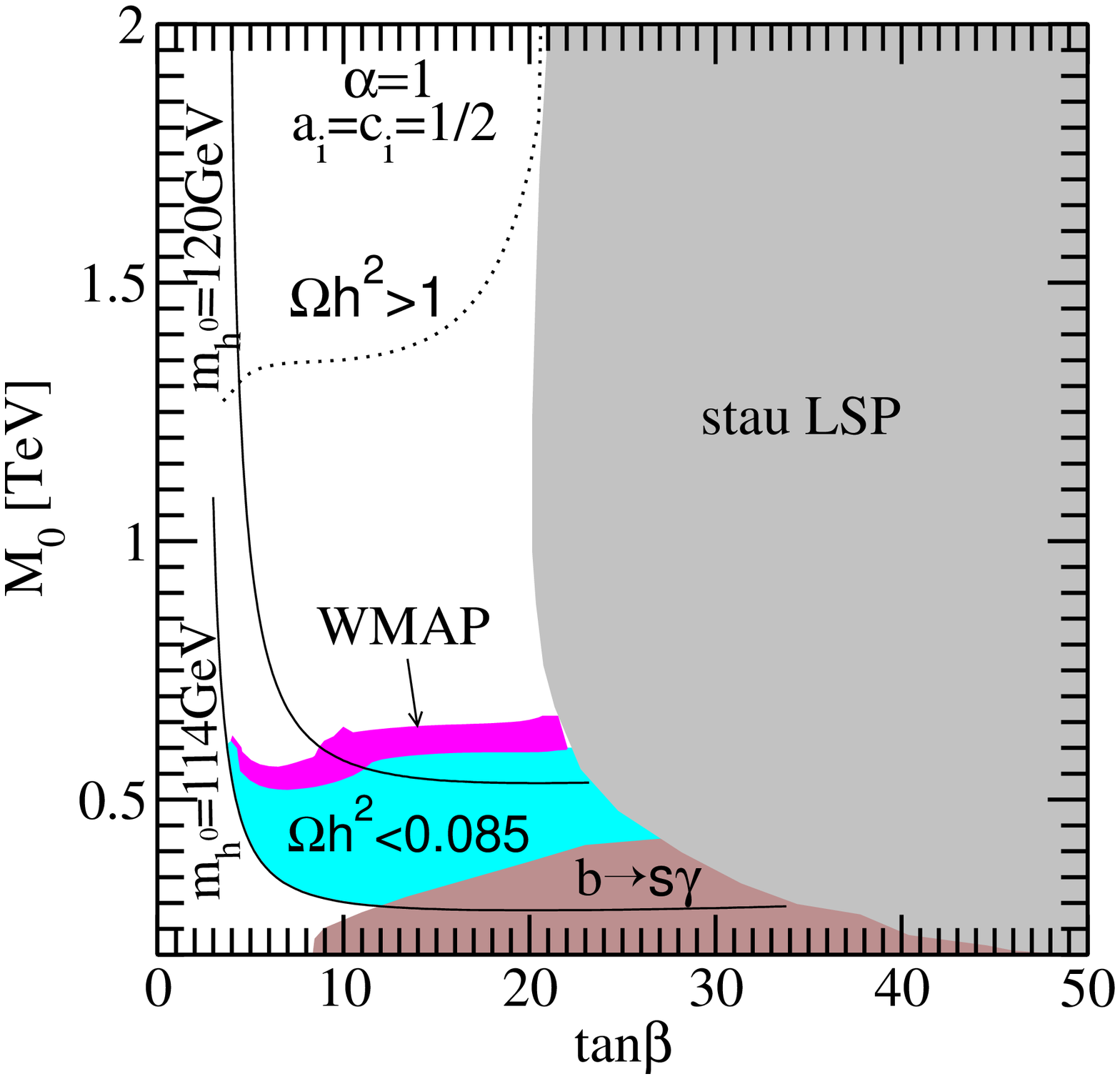}}
\end{tabular}
  \caption{The thermal relic abundance of the neutralino dark matter in the minimal case.
           \label{minimal-aboundance}}
\end{figure}

In figure \ref{minimal-aboundance}, we present the thermal relic abundance of the
neutralino in the minimal setup ($\alpha=1$) as a function of $\tan\beta$.
Left panel shows a case of universal modular weight $a_i=c_i=1$ 
 which corresponds to D7 matter fields in the KKLT setup
 \cite{Kachru:2003aw, Choi:2005uz}, while 
right panel shows alternative choice, $a_i=c_i=1/2$.
Within the magenta strip, the thermal abundance saturates
 the current WMAP bound, $0.085<\Omega h^2<0.119$ $(2\sigma)$
 and below it (cyan region)
 non-thermal contribution is required to fill the bound but allowed as well.
Green region is excluded due to stop LSP and gray region is excluded due
to stau LSP. Brown region is disfavored by $b \to s,\gamma$ branching ratio. We also eliminate a region in
 which the lightest Higgs mass is below the current SM Higgs mass bound.

In $a_i=c_i=1$ case, the lightest neutralino is almost bino whose
annihilation cross section is known to be small. However, annihilation 
 through heavy Higgs resonance ($\tan\beta \sim 22$) and coannihilation
 with stop (stau) near the boundary of stop (stau) LSP region reduce
 the relic abundance and a relatively heavy $M_0$ region satisfies the WMAP
 bound which evades the Higgs and $b \to s,\gamma$ bounds.
On the other hand, in $a_i=c_i=1/2$ case, stop LSP region disappears
 due to reduction of the RG contribution through the top Yukawa coupling. 
Reduction of the Higgs soft mass suppresses the heavy Higgs masses
 and removes the resonance away.
Instead, it introduces non-negligible higgsino
 components in the lightest neutralino.
This effect pushes up the WMAP region to $M_0 \approx 700$ GeV
 which is well above the Higgs and $b \to s,\gamma$ bounds.

\begin{figure}
\begin{tabular}{cc}
  \includegraphics[height=.24\textheight]{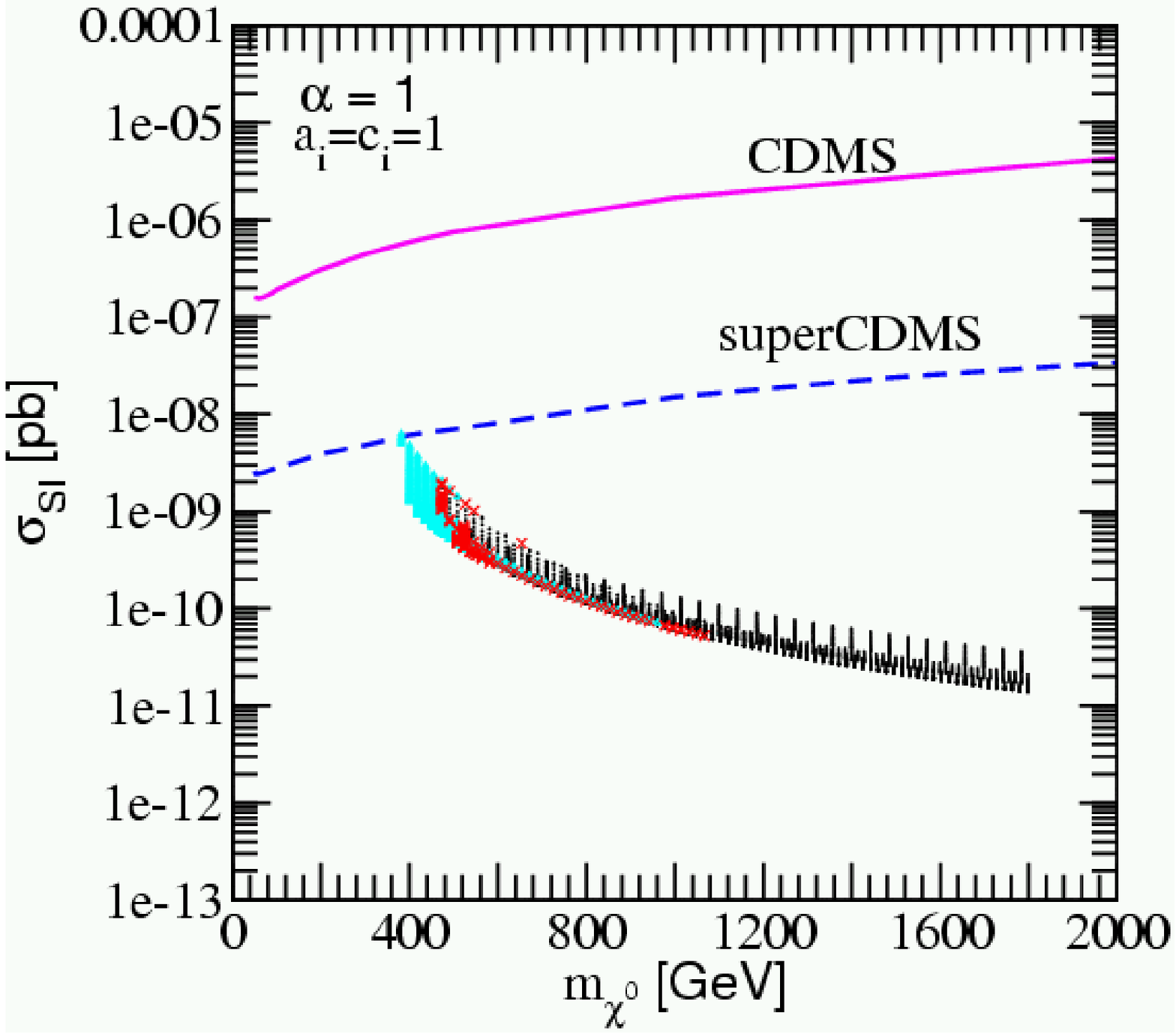} &
  \includegraphics[height=.24\textheight]{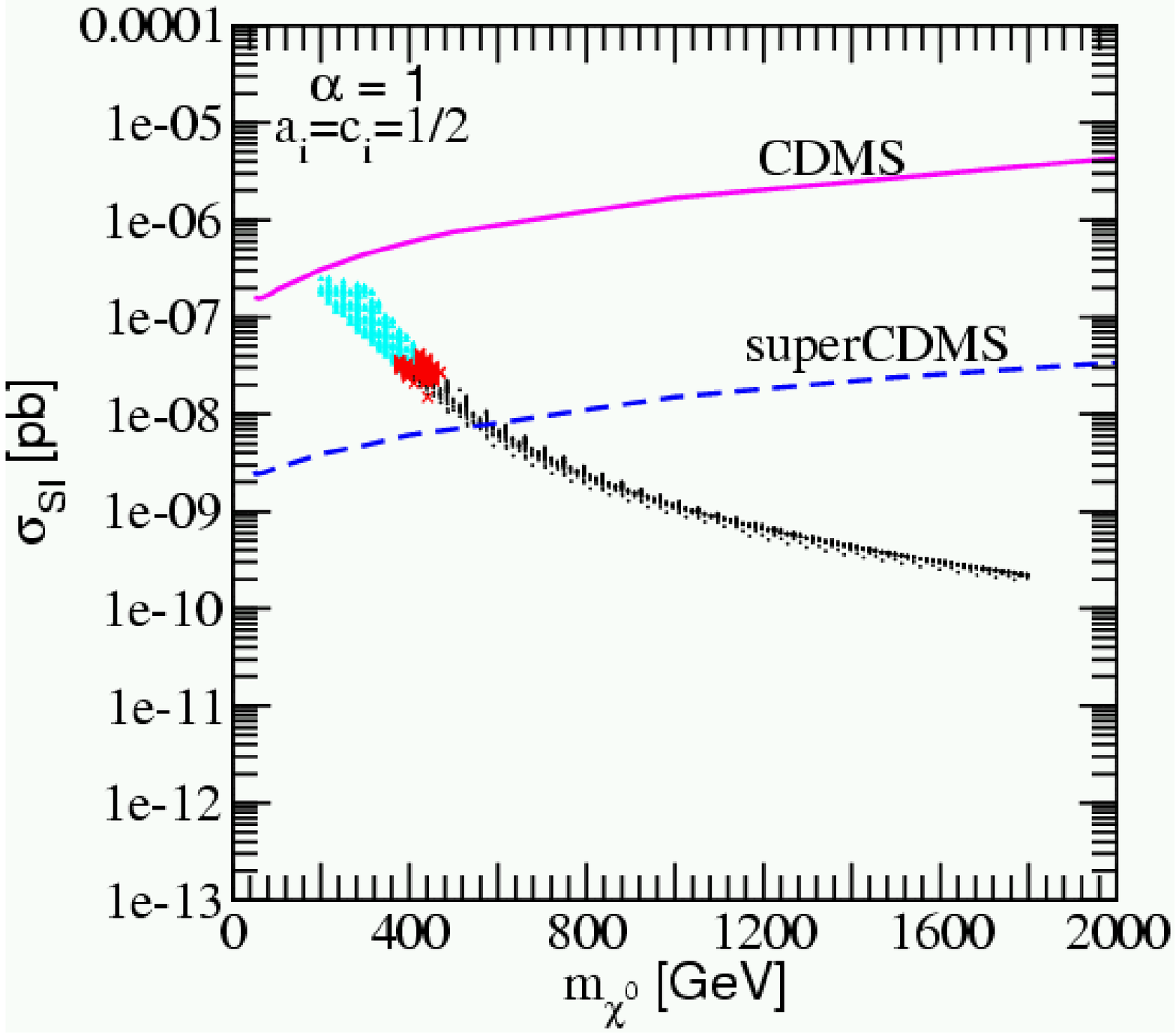} \\
  \includegraphics[height=.24\textheight]{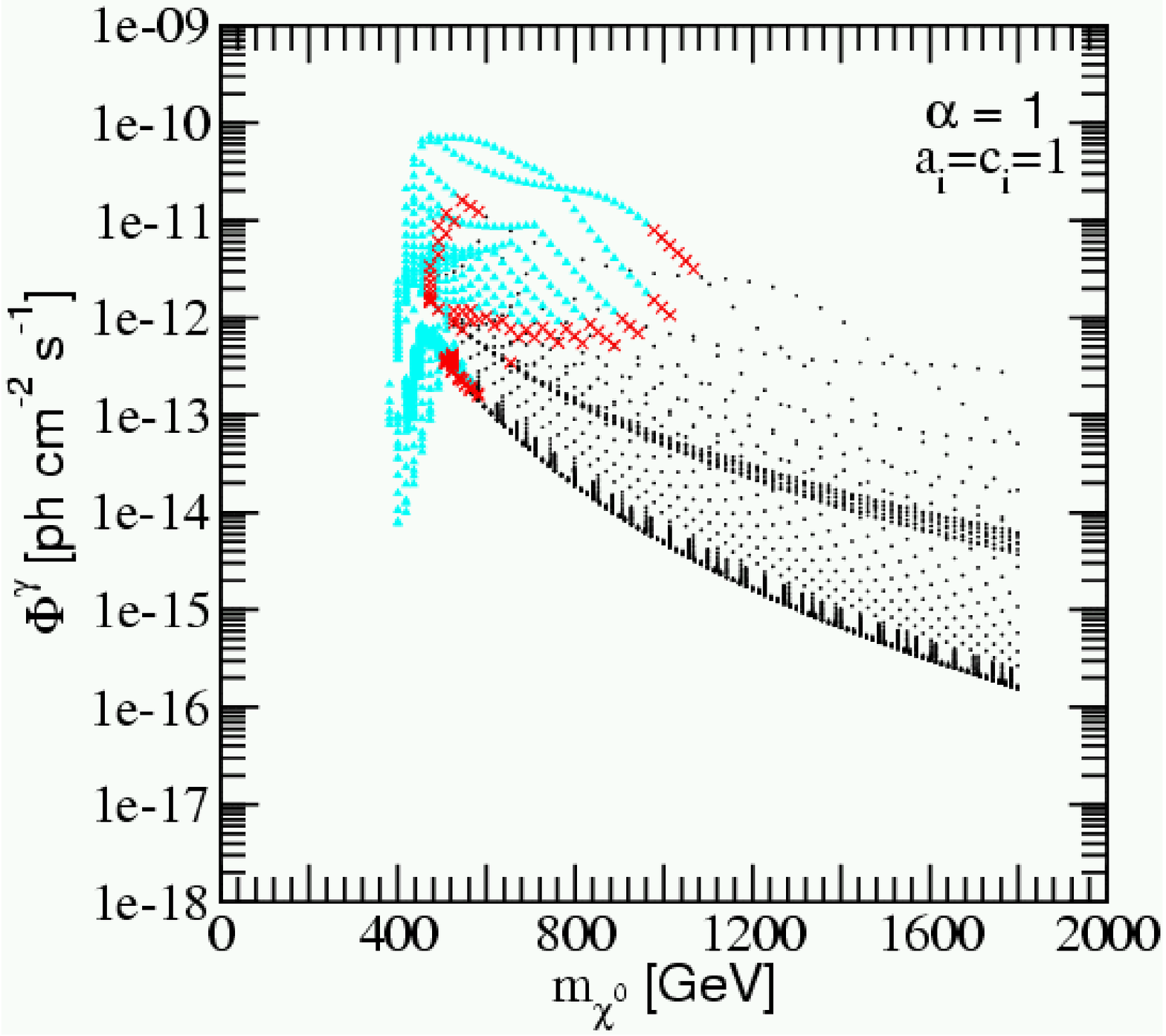} &
  \includegraphics[height=.24\textheight]{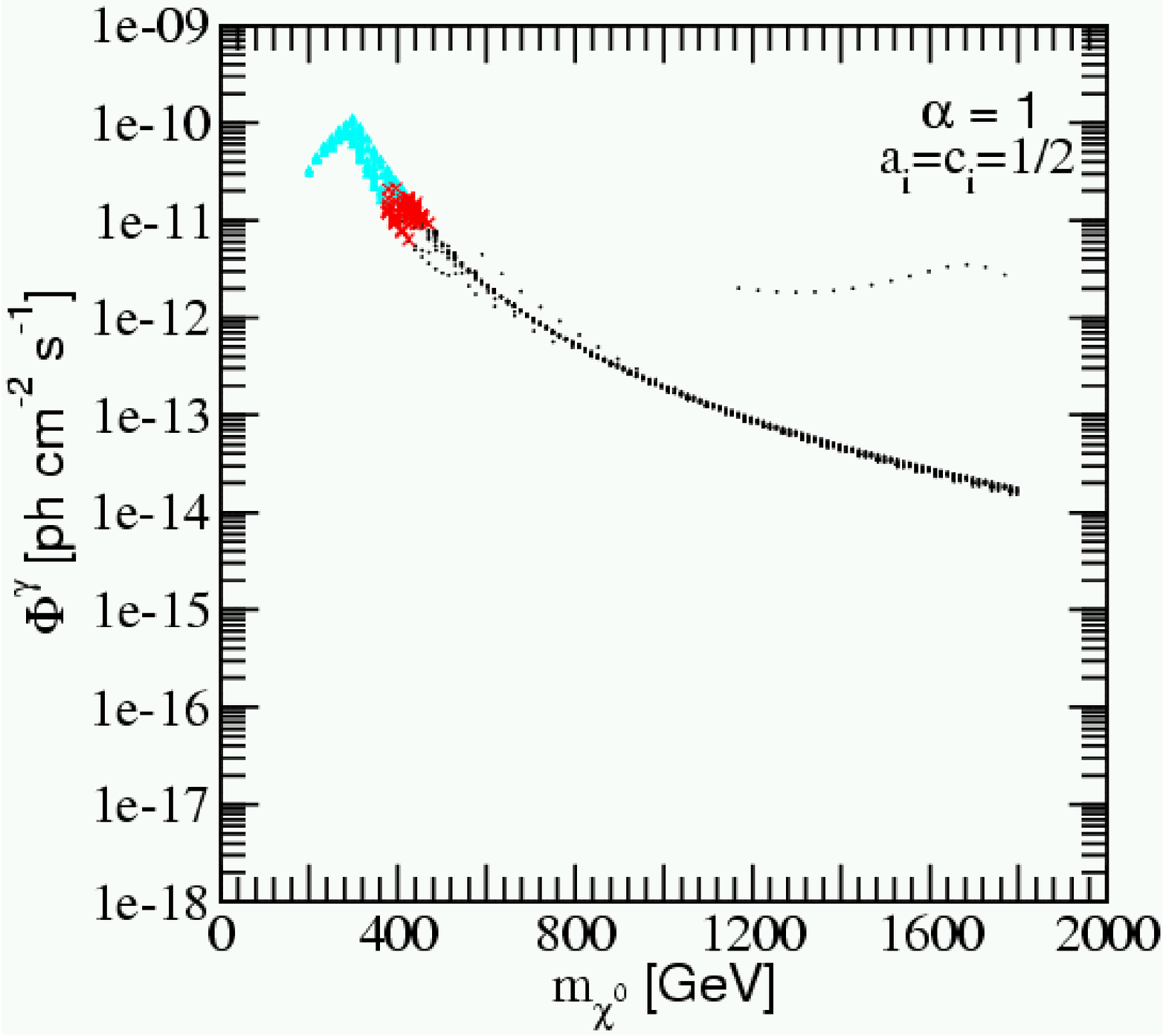}
\end{tabular}
  \caption{Prospects of detection of the neutralino dark matter in the
 minimal case.\label{minimal-crossection}}
\end{figure}

In the upper half of figure \ref{minimal-crossection}, we show the 
 spin-independent cross section of neutralino-proton scattering
 for the above two cases.
The red points saturate the WMAP bound and the cyan points fall
 below it.
Currently, CDMS experiment excludes the region above $\sim 10^{-6}$ pb
 and planned SuperCDMS is expected to reach sensitivity of $\sim
 10^{-9}$ pb \cite{supercdms}.
The spin-independent scattering proceeds through t-channel Higgs
 exchange and s-channel squark exchange.
 In many cases, the former dominates
 the process. In $a_i=c_i=1$ case, this Higgs exchange is
 suppressed due to suppressed higgsino components.
In the figure, all the allowed points are below the sensitivity of the planned direct detection experiment.
On the other hand, in $a_i=c_i=1/2$ case, all the WMAP allowed region
 (cyan) is above the SuperCDMS sensitivity due to the enhanced higgsino components.

In the lower half of figure \ref{minimal-crossection}, we show a prospect of indirect detection
 of neutralino dark matter through continuum gamma ray flux
 from the galactic center.
Again, the red points saturate the WMAP bound and cyan points fall below it.
Gamma ray flux from the galactic center depends not only on microscopic
 physics but also on poorly known galactic halo density profile. 
 Its integrated effect on the gamma ray flux is
 parameterized by the quantity, $\bar{J}(\Delta \Omega)$ which measures
 cuspiness of the profile over a spherical region
 of solid angle $\Delta \Omega$. In this paper, we use a conservative model
 (isothermal halo density profile) which predicts $\bar{J}(\Delta \Omega) \sim
 30$ with the detector angular resolution $\Delta \Omega=10^{-3}$ sr and
 set $E_{thr}=1$ GeV for gamma ray energy threshold.  
 However, note that, for example, an extreme spiked halo model can predict
 $10^4$ times enhanced $\bar{J}(\Delta \Omega)$ and gamma ray flux \cite{moore}.
In both $a_i=c_i=1$ and $a_i=c_i=1/2$ cases,
 the flux reaches $\sim 10^{-10} {\rm cm}^{-2} {\rm s}^{-1}$ which is
 similar to the expected reach of GLAST satellite.
 In $a_i=c_i=1$ where the neutralino is bino like, 
 the upper bound of the gamma ray flux similar to $a_i=c_i=1/2$ is
 reached by enhanced annihilation via heavy Higgs resonance. 
If the density profile is described by the extreme model, most of the
 WMAP allowed region (cyan) is explored by the future indirect experiment. 

\begin{figure}
\begin{tabular}{c c}
  \includegraphics[height=.24\textheight]{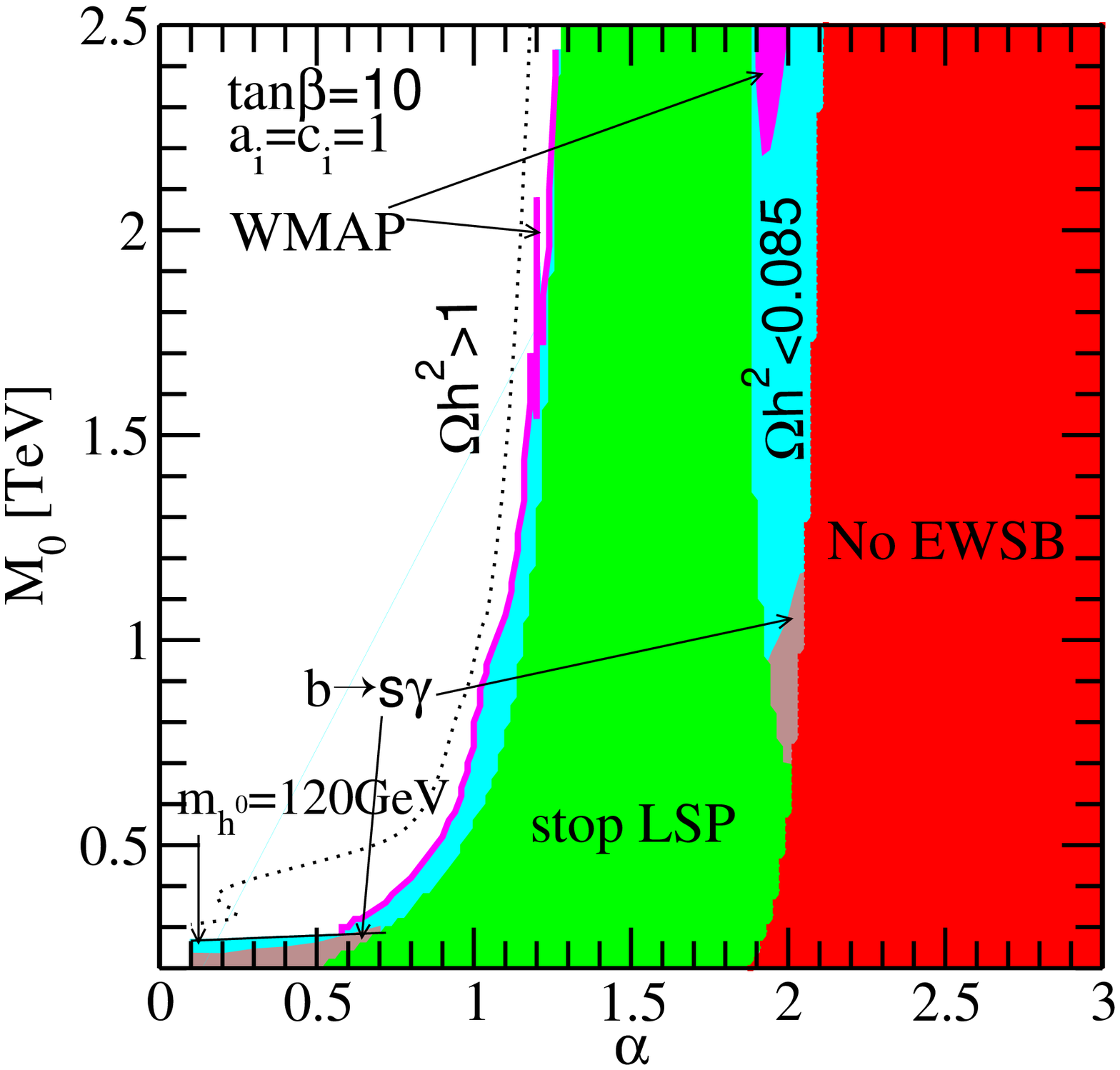} &
  \includegraphics[height=.24\textheight]{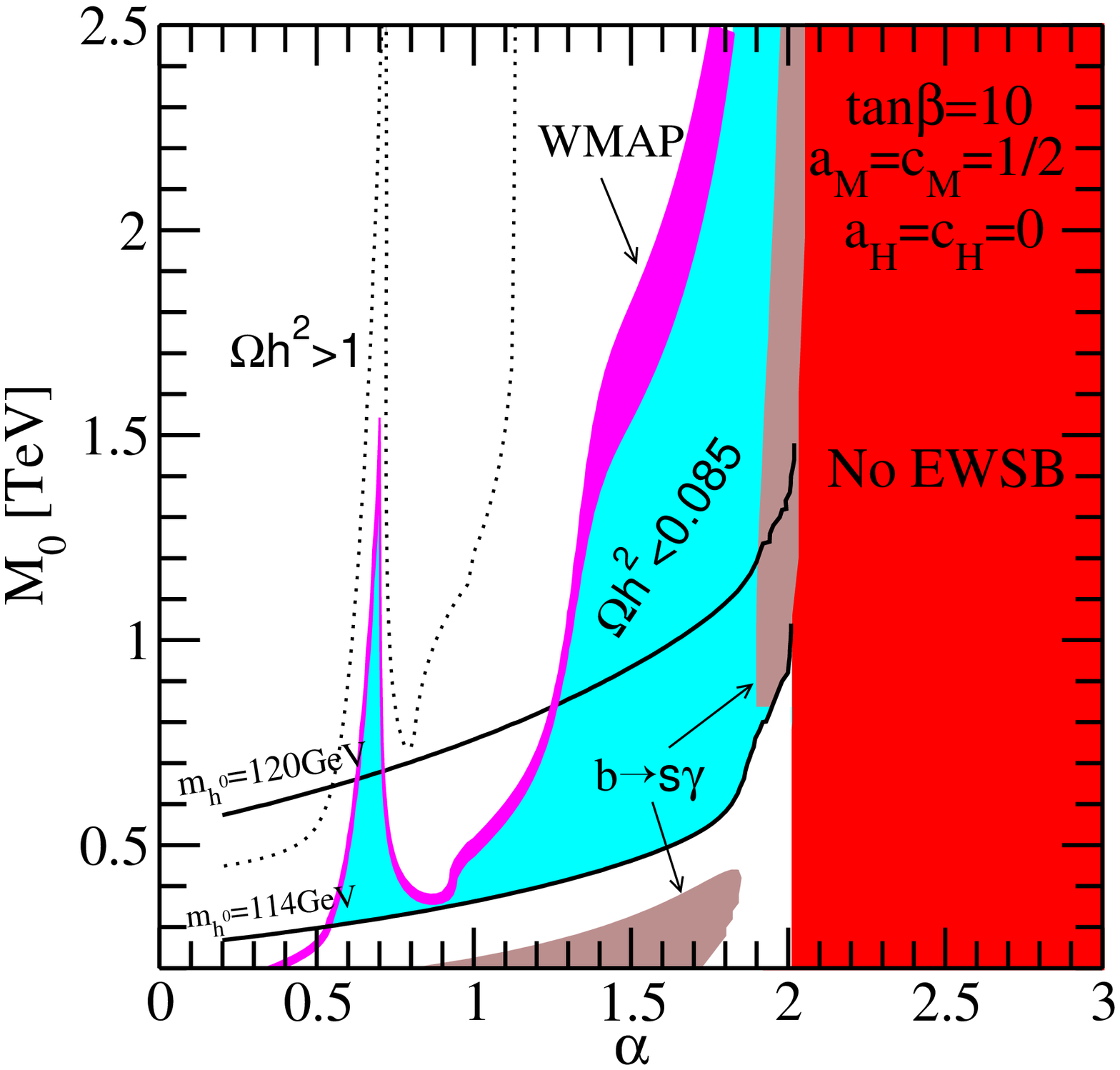}
\end{tabular}
  \caption{The thermal relic abundance of the neutralino dark matter in
 the non-minimal case.\label{non-minimal-aboundance}}
\end{figure}

Next we discuss the non-minimal case with $\alpha \neq 1$ which can be
 realized by dilaton-modulus mixing in gauge kinetic functions.
In figure \ref{non-minimal-aboundance}, we present thermal relic
 abundance as a function of $\alpha$ 
for universal modular weight $a_i=c_i=1$ which
 is a benchmark scenario in the KKLT model (left panel) and
 alternative case ($a_M=c_M=1/2$, $a_H=c_H=0$) where the matter and
 Higgs modular weights take different values (right panel).
The latter case is known to minimize fine-tuning in the EW
 symmetry breaking at $\alpha=2$ \cite{tevmirage}.
The red region is excluded due to no EW symmetry breaking.
The definition of allowed regions and other constraints are similar to those in figure \ref{minimal-aboundance}.

In the benchmark case, two allowed regions are separated by a stop LSP
 region. In the left hand side, neutralino is bino like as 
discussed in previous section. Relic abundance is suppressed 
 dominantly due to stop coannihilation for a particular choice of
 $\tan\beta$ (=10) in the figure.
In the narrow window around $\alpha=2$, the lightest
 neutralino is higgsino like and enhanced annihilation rate pushes up
 $M_0$ which saturates the WMAP result as heavy as 2.2 TeV.
In the non-universal case, the stop LSP region disappears
 due to the reduced RG contribution through the top Yukawa coupling.
The nature of LSP changes from pure bino to pure higgsino
 via bino-higgsino mixed region if we increase $\alpha$ from 0 to 2.
The heavy Higgs resonance appears around $\alpha=0.7$ for this particular
 choice of $\tan\beta$.
It is clearly observed that enhanced higgsino components due to lowered 
 mirage messenger scale expands the WMAP allowed region (cyan) relative to
 the conventional pure modulus mediation at the unification scale ($\alpha=0$).

\begin{figure}
\begin{tabular}{cc}
  \includegraphics[height=.24\textheight]{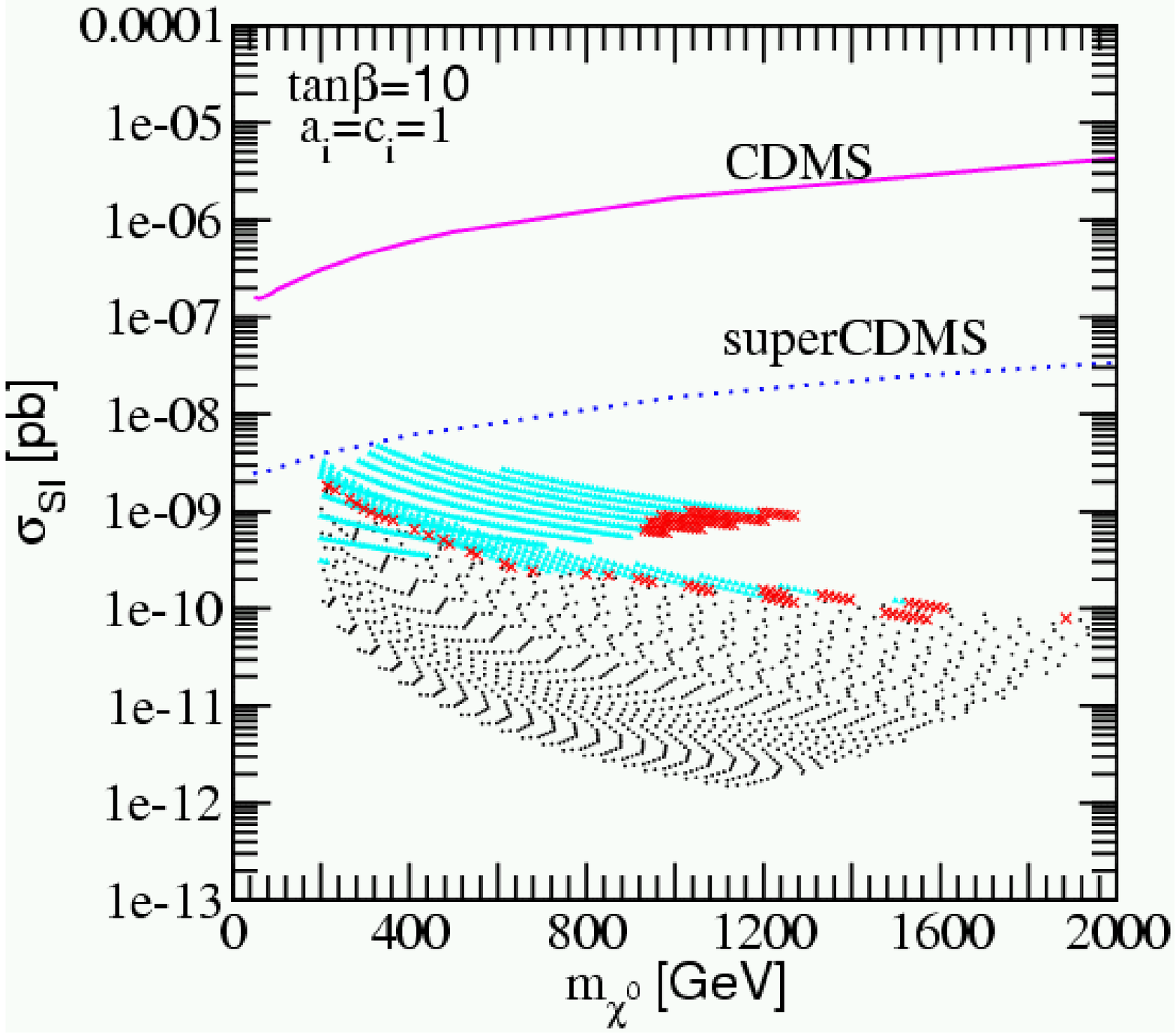} &
  \includegraphics[height=.24\textheight]{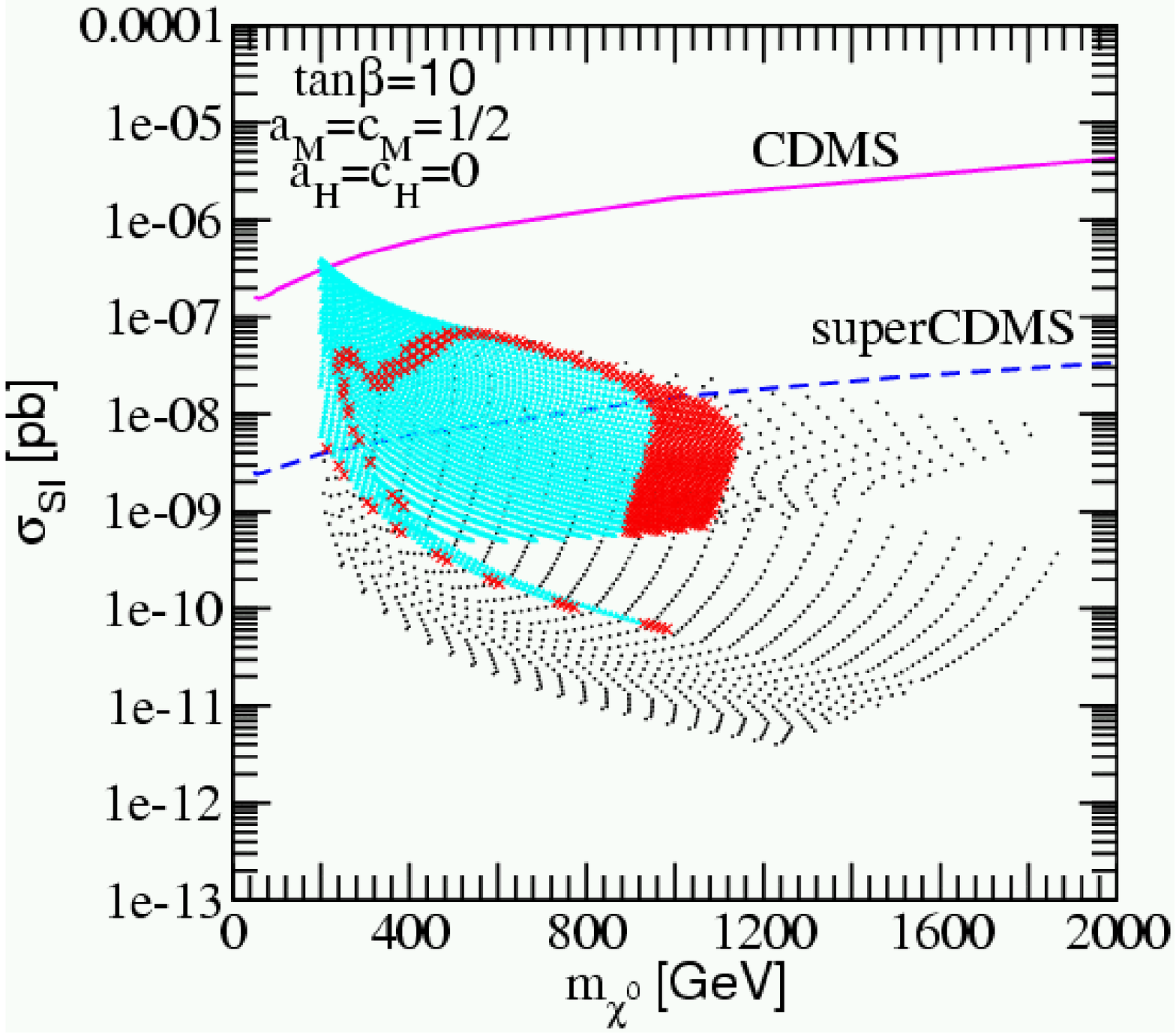} \\ 
  \includegraphics[height=.24\textheight]{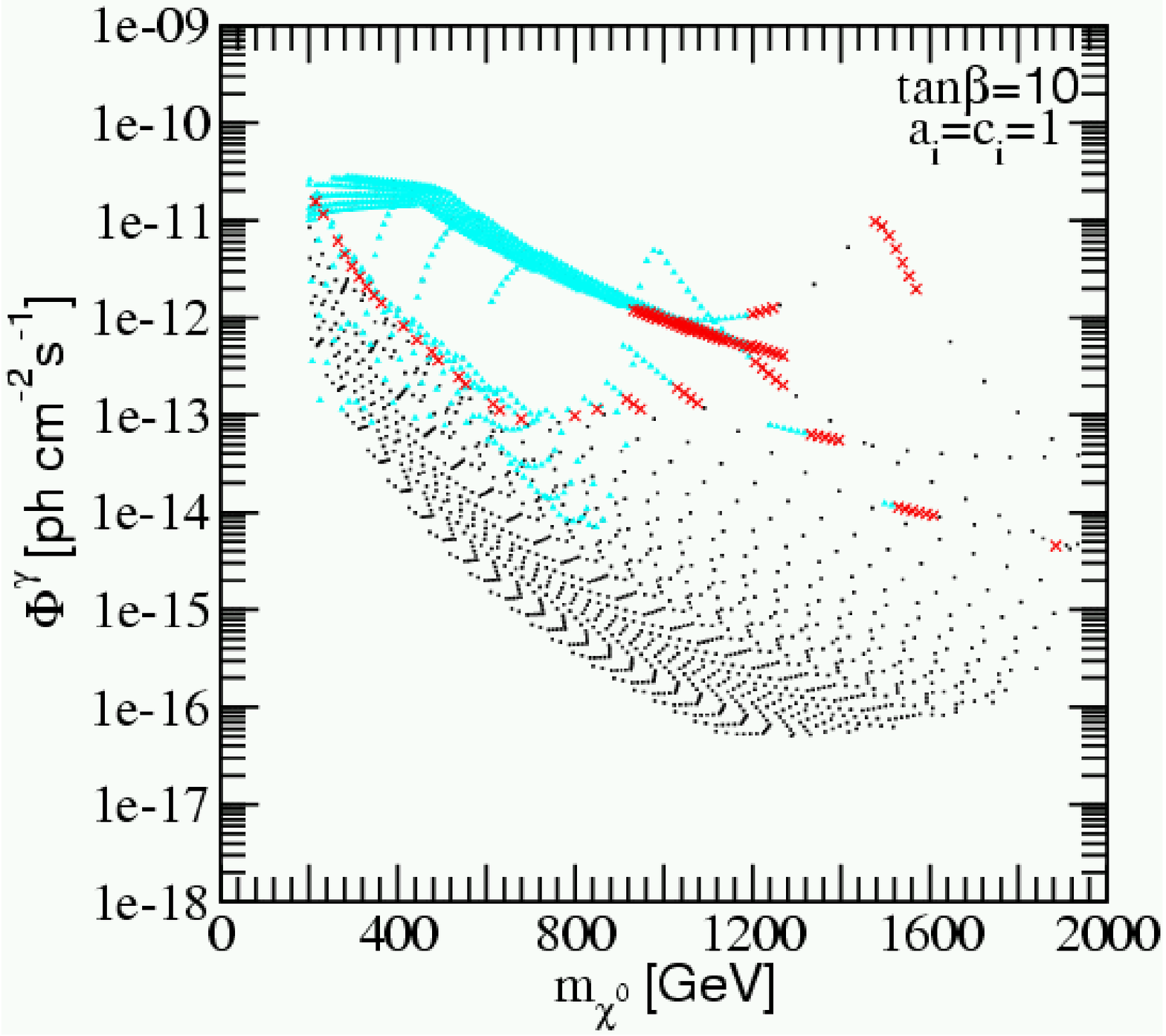} &
  \includegraphics[height=.24\textheight]{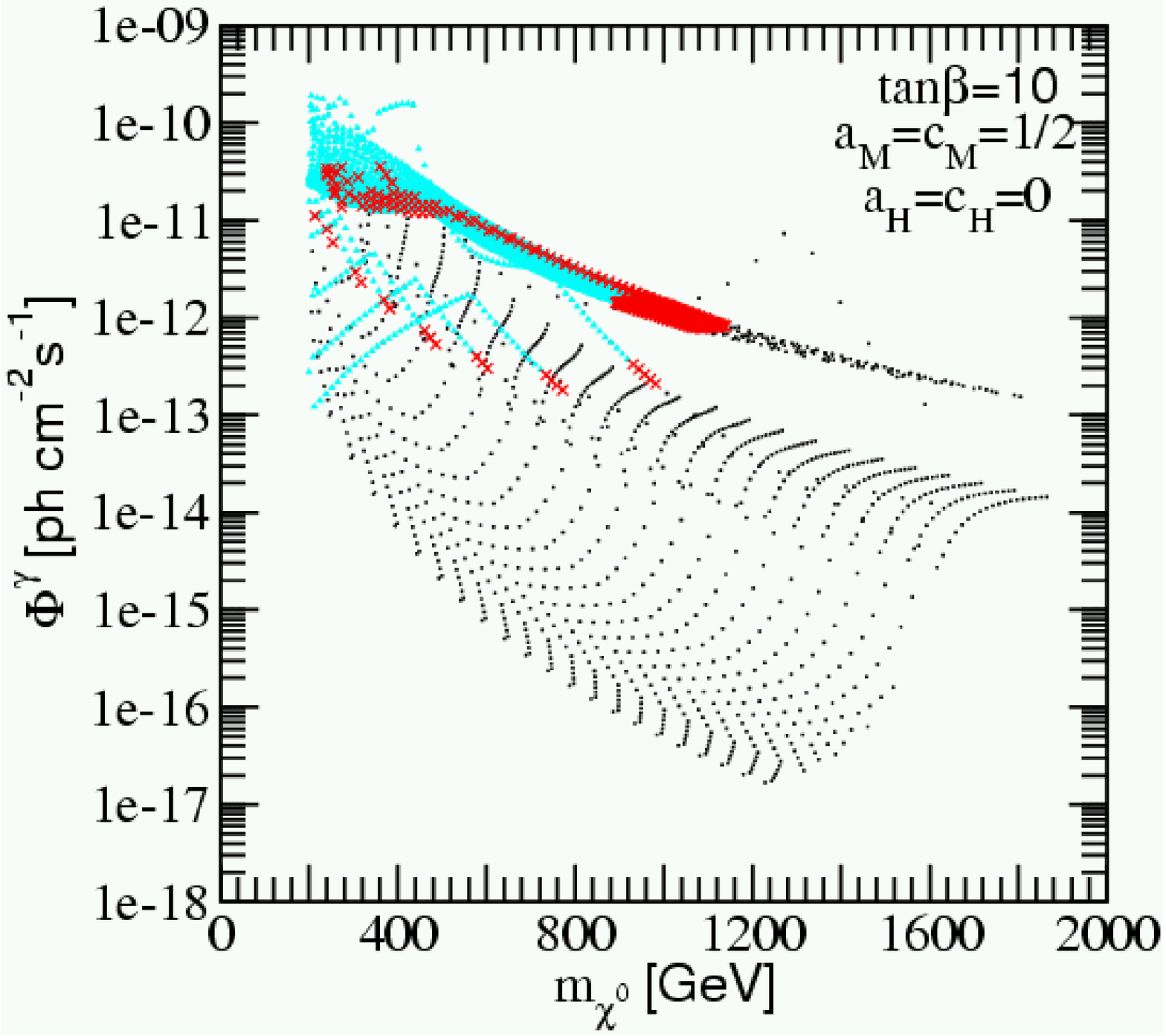} 
\end{tabular}
  \caption{Prospects of detection of the neutralino dark matter in the
 non-minimal case.\label{non-minimal-detection}}
\end{figure}

In the upper half of figure \ref{non-minimal-detection}, we show the
 spin-independent cross section of neutralino-proton scattering
 for the non-minimal case.
We scan $\alpha$ as in figure \ref{non-minimal-aboundance}
 for fixed $\tan\beta=10$. In the benchmark case, we can clearly 
discriminate two separate regions corresponding to the bino like and
 higgsino like regions. Because the t-channel Higgs exchange 
 is maximized in the mixed bino-higgsino region which is excluded due to stop
 LSP in this case, the upper bound of the cross section is below
 the expected reach of SuperCDMS.  On the other hand, in the alternative
 case, the mixed bino-higgsino region is available and the upper bound of
 the cross section even exceeds the current CDMS bound.

In the lower half of figure \ref{non-minimal-detection}, we present
 the continuum gamma ray flux from the galactic center for the
 non-minimal case. Parameters are scanned as the same as the upper half of the figure.
In the benchmark case, the flux reaches $\rm 2\times 10^{-11} cm^{-1}
s^{-1}$ which barely touches the expected reach of H.E.S.S. experiment \cite{hess}.
In the alternative case, enhanced annihilation due to the mixed
 bino-higgsino region lifts the upper bound to $\rm 2\times 10^{-10} cm^{-1}
s^{-1}$ which is similar to the expected reach of GLAST.
Note that these results are estimated using the conservative halo
 density profile and the extreme halo model can enhance the flux by $\sim 10^{-4}$.
In such a case, these indirect detection experiments can explore almost
entire WMAP allowed region.

\section{Conclusion}

In this paper, we briefly summarized the results of our analysis
 on thermal relic abundance and
 future prospects of direct and indirect detections
 of neutralino dark matter in mirage mediation.
The abundance is reduced due to stop/stau coannihilation, 
heavy Higgs resonance and enhanced higgsino components.
The WMAP bound can
 be satisfied in bulk of the parameter space.
Future direct experiments can detect
 mixed bino-higgsino dark matter which appears in non-minimal modular
 weight cases.
Indirect detection can explore
 the scenario if the galactic halo density has cuspy profile.


\begin{theacknowledgments}
This work is completed under the grant-in-aid for
scientific research on priority areas (No. 441): "Progress in
elementary particle physics of the 21 century through discoveries
of Higgs boson and supersymmetry" (No. 16081209) from the Ministry
of Education, Culture, Sports, Science and Technology of Japan.
Most of the numerical calculations were carried out on Altix3700 BX2
 at YITP in Kyoto University.
\end{theacknowledgments}

\end{document}